\documentstyle[12pt]{article}
\global\parskip 6pt

\normalsize

\let\oldtheequation=\theequation
\def\doteqs#1{\setcounter{equation}{0}
            \def\theequation{{#1}.\oldtheequation}}

\newcounter{sxn}
\def\sx#1{\addtocounter{sxn}{1} \bigskip\medskip \goodbreak
\noindent{\large\bf
\centerline{\thesxn.~~#1}} \nobreak \medskip}
\def\sxn#1{\sx{#1} \doteqs{\thesxn}}

\newcounter{axn}

\def\br{\begin{thebibliography}{abc}}
\def\er{\end{thebibliography}}

\def\be{\begin{equation}}
\def\ee{\end{equation}}
\def\bea{\begin{eqnarray}}
\def\eea{\end{eqnarray}}

\begin{document}
\begin{flushright}
\hfill{SINP-TNP/02-13}\\
\end{flushright}
\vspace*{1cm}
\thispagestyle{empty}
\centerline{\large\bf Near-Horizon Conformal Structure and }
\centerline{\large\bf Entropy of Schwarzschild Black
Holes \footnote{Expanded version of the talk presented by K.S.Gupta at the
Indo-Russian International Workshop on Quantum Gravity, Strings and
Integrable Models held at IMSc., Chennai, India during January 15-19,
2002.}}
\bigskip
\begin{center}
Kumar S. Gupta\footnote{Email: gupta@theory.saha.ernet.in}
\end{center}
\begin{center}
{\em Theory Division\\ 
Saha Institute of Nuclear Physics\\
1/AF Bidhannagar\\
Calcutta - 700 064,  India}\\
\vspace*{.5cm}
\end{center}
\vskip .2cm

\begin{abstract}

Near-horizon conformal structure of a massive Schwarzschild black hole of
mass $M$ is analyzed using a scalar field as a simple probe of the 
background geometry. The near-horizon dynamics is governed by an operator
which is related to the Virasoro algebra and admits a one-parameter family of 
self-adjoint extensions described by a real parameter $z$. When $z$ satisfies
a suitable contraint, the corresponding wavefunctions exhibit scaling 
behaviour in a band-like region near the horizon of the black hole. 
This formalism is consistent with the Bekenstein-Hawking entropy formula and
naturally produces the $- \frac{3}{2} {\rm log} M^2$ correction term to the
black hole entropy with other subleading corrections exponentially suppressed. 
This precise form for the black hole entropy is expected on general grounds in
any conformal field theoretic description of the problem. The presence of the 
Virasoro algebra and the scaling properties of the associated wavefunctions 
in the near-horizon region together with the appearance of the logarithmic 
correction to the Bekenstein-Hawking entropy provide strong evidence 
for the near-horizon conformal structure in this system.
\end{abstract}
\vspace*{.3cm}
\begin{center}
April 2002
\end{center}
\vspace*{1.0cm}
\baselineskip=14 pt
\newpage
\sxn{Introduction} 

The discovery of a conformal symmetry in the near-horizon region of a
black hole has led to interesting developments in quantum gravity
\cite{strom,carlip,solo}. By imposing a suitable boundary condition at the 
black hole horizon, it has been shown that the 
algebra of surface deformations contains a Virasoro subalgebra
\cite{carlip,solo,ghosh}. This approach
is based on extension of the idea of Brown and Henneaux \cite{brown} regarding
asymptotic symmetries of three-dimensional anti-de-Sitter gravity and is
essentially classical in nature. The knowledge of the Virasoro algebra in the 
near-horizon region provides a method to calculate the black hole entropy 
\cite{strom,carlip} and is indicative of the holographic nature of 
the system \cite{thorn,hooft1,suss,hooft}. The boundary conditions satisfied 
by the metric components at the horizon encode the appropriate physical
requirements for holography.

In this paper we provide an alternate approach for analyzing the
near-horizon conformal symmetry of black holes. Our method uses 
a massless scalar field coupled to the black hole background as a  
probe of the near-horizon geometry. It is useful to choose the probe in such
a way so that unnecessary complications do not obscure the essential 
physical features of the system. We shall initially restrict our attention
to massive Schwarzschild black holes and generalization to other geometries 
would be discussed later. With this in mind we consider the 
time-independent, zero angular momentum modes of a massless scalar field 
as a simple probe of the near-horizon geometry. The Klein-Gordon operator $H$ 
governing the dynamics of the probe in the near-horizon region
contains an inverse square potential term 
\cite{trg}. Systems with such a potential have been known to be associated 
with conformal mechanics \cite{fubini,vasiliev,town1,town2}. 
We shall show below that indeed the full Virasoro algebra arises naturally in
this system. $H$ however is not a element of the Virasoro algebra but 
belongs to the corresponding enveloping algebra \cite{ksg1}. This result is
obtained using a factorization of the operator $H$ and the 
inverse square term plays a crucial role in this algebraic analysis. 

	The quantum properties of the near-horizon KG operator $H$ yields
further valuable information about the system. In the quantum theory, 
$H$ admits a one-parameter family of self-adjoint extensions labelled by 
$e^{i z}$ where $z$ is a real number \cite{narn,reed}. For a generic
value of $z$, the system admits an infinite number of bound states. However,
it is only when $z$ is positive and satisfies the consistency condition
$z \sim 0$, 
that the bound states exhibit a scaling behaviour in the near-horizon 
region of the black hole. This property of the bound states reflects the 
existence of an underlying conformal structure in the near-horizon region. 
The self-adjoint parameter $z$ labels the domains of the Hamiltonian and is
therefore directly related to the boundary conditions satisfied by $H$.
The contraint on $z$ is thus conceptually analogous to the boundary conditions
satisfied by the metric components in Refs. \cite{carlip,solo} required for 
holography.
	
The spectrum of the operator $H$ is determined by boundary conditions which 
are encoded in the choice of the corresponding domain. At an operator level, 
the properties of the Hamiltonian are thus determined by the parameter $z$. 
On the other hand, the only physical input in the problem is the mass
$M$ of the black hole which must also play a role in determining
the spectrum. The parameters $z$ and $M$ thus play a
conceptually similar role and they are likely to be related to each other.
Establishing a relation between $z$ and $M$, which is consistent with the 
constraints of our model, is a crucial step in our approach. 

	The relationship between $z$ and $M$ naturally leads to the
identification of black hole entropy within this formalism \cite{ksg2}.
Our analysis is based on the identification of the bound states described
above with the quantum excitations of the black hole resulting from the
capture of the scalar field probe. These excitations  
will eventually decay through the emission of Hawking radiation.
The corresponding density of
states of the black hole is usually a function of the mass $M$ \cite{hooft}.
On the other hand, the density of states in our algebraic formalism has a 
smooth expression in terms of the self-adjoint parameter $z$.
In view of the
relation between $z$ and $M$, the density of states written as a function
of $z$ can be re-expressed in terms of the variable $M$.
This process leads to the  
identification of the black hole entropy consistent with the constraints of
the model. The entropy so obtained
is given by the Bekenstein-Hawking term together with
a leading logarithmic correction whose coefficient is $- \frac{3}{2}$. The
other subleading corrections terms are found to be exponentially suppressed.
Such a logarithmic correction term to the Bekenstein-Hawking entropy 
with a $- \frac{3}{2}$ coefficient was first 
found in the quantum geometry formalism \cite{partha} and has subsequently
appeared in several other publications \cite{car1,trg1,das1,danny,partha1}.
In particular, using an exact convergent expansion for the partition of a  
number due to Rademacher \cite{rade}, rather than the asymptotic formula due
to Hardy and Ramanujan \cite{ram}, it was shown in Ref. \cite{danny} 
that within any conformal field theoretic description, the black hole
entropy can be expressed in a series
where the leading logarithmic correction to Bekenstein-Hawking term always
appears with a universal coefficient of $- \frac{3}{2}$ with the other    
subleading terms exponentially suppressed.
Here we show that the entropy of the Schwarzschild black hole contains
correction terms precisely of this structure, which provides a strong 
evidence for the underlying near-horizon conformal structure of the system.

The plan of this paper is as follows. In Section 2 we provide an algebraic
formulation of the near-horizon dynamics where in terms of operators
appearing in the factorization of $H$. We also show how the Virasoro algebra
appears in this context. Some aspects of the representation theory of
Virasoro algebra is discussed in Section 3. The particular representation to
which $H$ belongs is explicitly obtained there. Section 4 discusses the
near-horizon quantization in terms of the self-adjoint extensions of $H$ and
the scaling properties of the corresponding wavefunctions is discussed in
Section 5. The appearance of the Bekenstein-Hawking entropy together with
the universal $- \frac{3}{2}$ logarithmic correction term is discussed in
Section 6. The paper is concluded in Section 7 with some remarks and an
outlook.

\sxn{Algebraic Formulation of the Near-Horizon Dynamics}

In this section, we consider the case of a scalar field probing the
near-horizon geometry of a massive Schwarzschild black hole of mass $M$.
The metric for an asymptotically flat, spherically symmetric and static
black hole in 3+1 dimensions is given by
\be
ds^2 = - F(r) dt^2 + F^{- 1}(r) dr^2 + r^2 d {\Omega}^2 =
g_{ij}dx^{i}dx^{j}.
\ee
For Schwarzschild black hole $F(r) = \left ( 1 - \frac{2M}{r} \right )$.

The action for a massless scalar field in the above background can be
written as 
\be
{\cal S} = - \frac{1}{2} \int \sqrt{|g|} g^{ij}{\partial}_i \phi {\partial}_j
\phi.
\ee
We shall restrict the analysis to the spherically symmetric and 
time-independent modes of the scalar field. The Klein-Gordon
operator governing the
near-horizon dynamics of these modes can then be written as \cite{trg}
\begin{equation}
H = - {\frac{d^2}{dx^2}} + {a \over x^2},
\end{equation}
where $a$ is a real dimensionless constant, and 
$x = r - 2M $ is the
near-horizon coordinate. For the Schwarzschild background,
we have $a = - \frac{1}{4}$.
For the moment, however, we can consider a general value of $a$.

The starting point of our formulation is the observation that the 
operator $H$ can be factorized
as
\be
H = A_+A_-,
\label{H}
\ee  
where
\be
A_{\pm} = \pm \frac{d}{dx} + \frac{b}{x},
\ee
and 
\be
b =  \frac{1}{2} \pm \frac{\sqrt{1 + 4a}}{2}.
\ee 
We note that $a = - \frac{1}{4}$ is the minimum value of $a$
for which $b$ is
real. For real values of $b$, $A_+$ and $A_-$ are formal
adjoints of each other (with 
respect to the measure $dx$), and consequently $H$ is formally
a positive quantity
(there are some subtleties to this argument arising from the
self-adjoint extensions of $H$ which will be discussed later).
When $a < - \frac{1}{4}$, $b$ is no longer real and
$A_+$ and $A_-$ are not
even formal adjoints of each other. However,
$H$ can still be factorized as in Eqn. (\ref{H}),
but it is no longer a positive definite quantity.
It can still be
made self-adjoint \cite{case}, but remains unbounded
from below; this case has been analyzed in \cite{ksg}.

Let us now define the operators
\bea
L_n &=& -x^{n + 1} \frac{d}{dx},~~~~~~~~~n \in {\mathbf Z},\\
P_m &=& \frac{1}{x^m},~~~~~~~~~m \in {\mathbf Z}.
\eea
In terms of these operators, $A_{\pm}$ and  $H$ can be written as
\bea
A_{\pm} &=& \mp L_{-1} + b P_1,\\
H &=& (- L_{-1} + b P_1) (L_{-1} + b P_1).
\eea
Thus,
$L_{-1}$ and $P_1$ are the  basic operators appearing in the factorization
of $H$. Taking all possible commutators of these operators between
themselves and with $H$, we obtain the following relations
\begin{eqnarray} {}
[ P_{m}, P_{n} ] &=& 0,\\ {}
[ L_{m}, P_{n} ] &=& n P_{n - m},\\ {}
[ L_{m}, L_{n} ] &=& ( m - n) L_{m + n} + \frac{c}{12}
(m^{3} - m)\delta_{m+n, 0},\\ {}
[P_m, H] &=& m (m+1) P_{m + 2} + 2 m L_{-m -2}, \\{}
[L_m, H] &=& 2b(b-1)P_{2-m} - (m+1)(L_{-1}L_{m-1} + L_{m-1}L_{-1}).
\end{eqnarray}
Eqn. (2.13) describes a Virasoro algebra with central charge $c$.
Note that the algebra of the generators defined  in Eqn. (2.7)
would lead to
$[ L_{m}, L_{n} ] = ( m - n) L_{m + n}$. However, this algebra
is known to
admit a non-trivial central extension. Moreover, for any
irreducible unitary highest weight representation of
this algebra,
$c \neq 0$ \cite{godd}. For these reasons,
we have included the central term explicitly
in Eqn. (2.13).

Eqns. (2.11 - 2.13) describe the semidirect product of the Virasoro algebra
with an abelian algebra satisfied by the shift operators $\{P_{m}\}$.
Henceforth, we denote this semidirect product algebra by
${\cal {A}}$. 
Note that $L_{- 1}$ and $P_1$ are the only  generators that appear in $H$.
Starting with these two generators, and using Eqns. (2.14) and (2.15),
we see that the only operators which appear are the Virasoro generators
with negative index (except $L_{-2}$), and the shift
generators with positive index.
Thus, $L_m$ with $m \geq 0$ and $P_m$
with $m \leq 0$ do not appear in the above expressions. In the next
section, we will discuss how these quantities are generated.

Although the  algebra of Virasoro and shift generators
has a semidirect product structure, the operator $H$
however does not belong to this algebra.
This is due to the fact that the right-hand side of Eqn. (2.15)
contains products of the Virasoro generators. While
such products are not elements of the algebra, they do
belong to the corresponding enveloping algebra.
The given system is thus seen to be described by the 
enveloping algebra of the Virasoro generators, together with the abelian
algebra of the shift operators. This algebraic system has been extensively
studied in the literature \cite{kac}.

\sxn {Representation}

We shall now discuss the representation theory of the algebra
${\cal{A}}$, and the implications for the quantum properties of the
operator $H$. 
The eigenvalue equation of interest is
\be
H | \psi \rangle =  {\cal E} | \psi \rangle,
\ee
with the boundary condition that $\psi (0) = 0$. 
We are especially
interested in the bound state sector of $H$.  
As we have seen, the operator $H$ can be expressed in terms
of certain operators that belong to the
algebra ${\cal {A}}$. This observation allows us to
give a description of the  states of $H$ in terms of the
representation spaces of ${\cal {A}}$.
We first recall the relevant aspects of the representation theory
of ${\cal{A}}$.

Following \cite{kac},
we introduce the space $V_{\alpha, \beta}$ of densities
containing elements of the form
$P(x) x^\alpha (dx)^\beta$, where $\alpha, \beta$ are complex
numbers, in general.
Here, $P(x)$ is an arbitrary polynomial in $x$ and $x^{-1}$,
where $x$  is now treated as a complex variable.
It may be noted that the algebra ${\cal {A}}$ remains unchanged even
when $x$ is complex. It is known that
$V_{\alpha, \beta}$ carries a representation of the algebra ${\cal {A}}$.
The space $V_{\alpha, \beta}$ is spanned by a set of vectors,
$\omega_m = x^{m + \alpha} (dx)^\beta$, where $m \in {\mathbf Z}$.
The Virasoro generators and the shift operators have the following
action
on the basis vectors $\omega_m$,
\bea
P_n (\omega_m) &=& \omega_{m-n},\\
L_n (\omega_m) &=& - (m + \alpha + \beta + n \beta) \omega_{n + m}.
\eea 
The representation $V_{\alpha, \beta}$ is reducible
if  $\alpha \in {\mathbf Z}$ and if $\beta = 0$ or $1$;
otherwise it is irreducible.

The requirement of unitarity of the representation $V_{\alpha, \beta}$
leads to several important consequences. In any unitary representation of 
${\cal {A}}$, the Virasoro generators must satisfy the condition
${L_{- m}^{\dagger}} = L_m$. In the previous section, we saw
that
$L_{-2}$ and $L_m$ for $m \geq 0$ did not appear in the algebraic
structure generated by the basic operators appearing in
the factorization of $H$.
However, the requirement of a unitary
representation now leads to the inclusion of $L_m$ for
$m > 0$. The remaining generators now appear through
appropriate commutators, thus completing the algebra ${\cal A}$.

Unitarity also constrains the parameters $\alpha$ and
$\beta$, which must now satisfy the conditions
\bea
\beta + \bar{\beta} &=& 1,\\
\alpha + \beta &=& {\bar{\alpha}} + {\bar{\beta}},
\eea
where ${\bar{\alpha}}$ denotes the complex conjugate of $\alpha$.
Finally, the central charge $c$ in the representation $V_{\alpha, \beta}$
is given by
\be
c (\beta) = -12 \beta^2 + 12 \beta -2.
\ee

The above representation of ${\cal {A}}$ can now be used to analyze the
eigenvalue problem of Eqn. (3.1). We would like to have a series solution to
the differential Eqn. (3.1), and consequently choose
an ansatz for the  wave function $\mid\!\psi \rangle$ 
given by
\be
| \psi \rangle = \sum^{\infty}_{n=0} c_n \omega_n.
\ee     
Furthermore, the
operator $H$, as written in Eqn. (\ref{H}), has a well-defined action on
$| \psi \rangle $. From Eqn. (3.3), it may be seen that
\be
L_{-1} (\omega_n) = - (n + \alpha) \omega_{n-1},
\ee
which is independent of $\beta$. Therefore, it appears that an
eigenfunction of $H$ may be constructed from elements of
$V_{\alpha, \beta}$
for arbitrary  $\beta$. However, the unitarity conditions of
Eqns. (3.4-3.5) put
severe restrictions on $\beta$, as we shall see below.

The indicial equation 
obtained by substituting Eqn. (3.7) in Eqn. (3.1) gives
\be
\alpha = b,\;{\mathrm{or}}~(1 - b).
\ee
To proceed, we analyze the cases (i) $a \geq - \frac{1}{4}$,
and (ii) $ a < - \frac{1}{4}$ separately.

\noindent
(i) $a \geq - \frac{1}{4}$.

This is the main case of interest as it includes
the value of $a$ for the
Schwarzschild background.
It follows from Eqn. (2.4) and
Eqn. (3.9) that $b$ and $\alpha$ are real. The unitarity condition
of Eqn (3.4) now fixes the value of $\beta = \frac{1}{2}$, and the
corresponding central charge is given by $c = 1$. It may be noted that 
relation of the central charge calculated here to that appearing in
the calculation of black hole entropy depends on geometric properties
of the black hole in question. We do not address this issue here.
Thus, we see that for the Schwarzschild black hole, we have identified
the relevant representation space as $V_{1/2,1/2}$.

\noindent
(ii) $ a < - \frac{1}{4}$.

In this case, we can write $ a = - \frac{1}{4} - \mu^2$
where $\mu \in {\mathbf R}$.
It follows from Eqn. (2.4), that $b = \frac{1}{2} \pm i \mu$. Eqn. (3.9)
then gives $\alpha = \frac{1}{2} \pm i \mu$, or $-\frac{1}{2} \mp i \mu$.
Let us take the   
case when $\alpha = \frac{1}{2} + i \mu$, the other
cases being similar. From Eqns. (3.4) and (3.5), we find
$\beta = \frac{1}{2} - i \mu$.
The value of the corresponding central charge is given by 
$c = 1 + 12 \mu^2$. The operator $H$ in this case can be made self-adjoint
but its spectrum remains unbounded from below \cite{case,ksg}. The algebraic
description, however, always leads to a well-defined representation.

\sxn{Near-Horizon Quantization}

We return now to the eigenvalue problem for the
differential operator $H$, and focus attention on the Schwarzschild
background, for which $a = -\frac{1}{4}$.
As already mentioned, we are interested in the
bound state sector of $H$. The corresponding Scrodinger's equation can be
written as 
\be
H \psi = {\cal E} \psi, ~~~ \psi (0) = 0
\ee
where $\psi \in L^2[R^+, dr]$. 
$H$ appearing in the above equation is an example of an unbounded linear 
operator on a Hilbert space.
Below we shall first summarize some basic properties of these operators
which would be useful for our analysis.

        Let $T$ be an unbounded differential operator acting on a Hilbert
space ${\cal H}$ and let $(\gamma , \delta )$ denote the inner product
of the elements $\gamma , \delta \in {\cal H}$.
By the Hellinger-Toeplitz theorem \cite{reed}, $T$  has a well defined    
action only on a dense subset $D(T)$ of the Hilbert space  ${\cal H}$.    
$D(T)$ is known as the domain of the operator $T$. Let $D(T^*)$ be the set
of $\phi \in {\cal H}$ for which there is a unique $\eta \in {\cal H}$ with 
$(T \xi , \phi) = (\xi , \eta )~ \forall~ \xi \in D(T)$. For each such
$\phi \in D(T^*)$ we define $T^* \phi = \eta$. $T^*$ is called the adjoint  
of the operator $T$ and $D(T^*)$ is the corresponding domain of the adjoint.

The operator $T$ is called symmetric or Hermitian if $T \subset T^*$,     
i.e. if $D(T) \subset D(T^*)$ and $T \phi = T^* \phi~ \forall~ \phi \in   
D(T)$. Equivalently, $T$ is symmetric iff $(T \phi, \eta) = (\phi, T \eta)
~ \forall ~ \phi, \eta \in D(T)$. The operator $T$ is called self-adjoint
iff $T = T^*$ and $D(T) = D(T^*)$.

We now state the criterion to determine if a symmetric operator $T$ is
self-adjoint. For this purpose let us define the deficiency subspaces 
$K_{\pm} \equiv {\rm Ker}(i \mp T^*)$ and the 
deficiency indices $n_{\pm}(T) \equiv
{\rm dim} [K_{\pm}]$. $T$ is (essentially) self-adjoint iff
$( n_+ , n_- ) = (0,0)$.
$T$ has self-adjoint extensions iff $n_+ = n_-$. There is a one-to-one
correspondence between self-adjoint extensions of $T$ and unitary maps
from $K_+$ into $K_-$. Finally if $n_+ \neq n_-$, then $T$ has no
self-adjoint extensions.

We now return to the discussion of the operator $H$.
On a domain $ D(H) \equiv \{\phi (0) = \phi^{\prime} (0) = 0,~
\phi,~ \phi^{\prime}~  {\rm absolutely~ continuous} \} $, $H$ is a symmetric
operator with deficiency indices (1,1). The corresponding deficiency
subspaces $ K_{\pm} $ are 1-dimensional and are spanned by
\bea
\phi_+ (x) &=& x^{\frac{1}{2}}H^{(1)}_0 (xe^{i \frac{ \pi}{4}}),\\
\phi_- (x) &=& x^{\frac{1}{2}}H^{(2)}_0 (xe^{-i \frac{ \pi}{4}})  
\eea
respectively, where $ H^{(1)}_0 $ and $ H^{(2)}_0 $ are Hankel functions.
Thus, the operator $H$ is not self-adjoint on
$D(H)$ but admits a one-parameter family of self-adjoint extensions, 
labelled by 
unitary maps from $ K_+ $ into $ K_- $.  The self-adjoint
extensions of $H$ are thus labelled by $e^{i z}$ where $z \in R$. 
The operator $H$ is self-adjoint in the domain
$D_z(H)$ which contains all the elements of $D(H)$ together
with elements of the form $ \phi_+ (r) + e^{i z} \phi_- $.
Each value of the parameter $z$ defines a particular domain $D_z(H)$ on which 
$H$ is self-adjoint and thus corresponds to a particular choice of boundary
condition. 

For an arbitrary value of the self-adjoint parameter $z$,
the normalized bound state solutions of Eqn. (5.1) 
are given by
\be
\psi_n (x) =  \sqrt{2 E_n x} K_0\left( \sqrt{E_n} x \right )
\ee
where ${\cal E}_n = -E_n$. 
In order to find the energy, we use the fact that
if $H$ has to be self-adjoint, the eigenfunction
$\psi_n (x)$ must belong to the domain $D_z(H)$. This leads to the
expression of energy given by  
and 
\be
{\cal E}_n = -E_n = - {\exp}\left[\frac{\pi}{2} (1 - 8n) {\cot} 
\frac{z}{2}\right]
\ee
respectively, where $n$ is an integer
and $K_{0}$ is the modified Bessel function \cite{narn,trg}. Thus 
for each value of $z$, the operator $H$ admits an infinite number of negative 
energy solutions. In our formalism, these solutions are interpreted as bound
state excitations of the black hole due to the capture of the scalar field.
As is obvious from Eqns. (4.4) and (4.5), different choices of $z$ leads to
inequivalent quantization of the system. The physics of the system is thus
encoded in the choice of the parameter $z$.

\sxn{Scaling Properties}

As we have seen, the Virasoro algebra
plays an important role in determining the spectrum of $H$.
Since this operator is associated with a probe of the near-horizon
geometry, one might expect that the corresponding wave functions
would exhibit certain scaling behaviour in this
region.

Firstly,  let us recall that
the horizon in this picture is located at $x = 0$. However,
the wave functions
$\psi_n$ vanish at $x = 0$, and therefore do not exhibit
any non-trivial scaling.
Nevertheless, it is of interest to examine the behaviour of
the wave functions near the horizon. For $x \sim 0$,
the wave functions have the form
\be
\psi_n = \sqrt{2 E_n x}\left(B - \ln \left( \sqrt{E_n} x\right)\right),
\ee
where $B = \ln 2 - \gamma$, and $\gamma$ is Euler's constant \cite{abr}.
While the logarithmic term, in general, breaks the scaling property,
one notices that it vanishes at the  point
$x_0 \sim 1/\sqrt{E_n}$,
where the wave functions exhibit a scaling
behaviour. 
The entire analysis so far, including the existence of the Virasoro
algebra, is valid only in the near-horizon region of the black hole.
Therefore, consistency of the above scaling behaviour
requires that $x_0$  belongs to the near-horizon region.
The minimum value of $x_0$ is obtained when $E_n$ is maximum. When the
parameter $z$ appearing in the self-adjoint extension of $H$
is positive,
the maximum value of $E_n$ is given by 
\be 
E_0 = {\exp}\left[\frac{\pi}{2} {\cot} \frac{z}{2}\right].
\ee
However, when $z$ is negative, the maximum value of $E_n$ is obtained when
$n \rightarrow \infty$. In this case, $x_0 \rightarrow 0$ where, as we have
seen before, the wave function vanishes and scaling becomes trivial. We
therefore conclude that 
\be
x_0 \sim \frac{1}{\sqrt{E_0}},~ z > 0
\ee
is the minimum value of $x_0$.
It remains to show that $x_0$ given by Eqn. (5.3) belongs to
the near-horizon region. We first note that we are free
to set $z$ to an arbitrary positive value.
Thus, we consider  $z >0 $ such that
${\cot} \frac{z}{2} >> 1$; this is achieved by choosing
$z \sim 0$.
For all such $z$,  we find that $x_0$ is small but nonzero, and
thus belongs to the near-horizon region.
In effect, we can use the freedom in the choice of
$z$ to restrict $x_0$ to the near-horizon region.

We now consider a band-like region
$\Delta = [x_{0} -\delta/\sqrt{E_{0}}, x_{0} + \delta/\sqrt{E_{0}}]$,
where $\delta \sim 0$ is real and positive. The region $\Delta$ thus belongs
to the near-horizon region of the black hole.
At a point $x$ in the region $\Delta$,
the leading behaviour of $\psi_n$ is given by 
\be
\psi_n = \sqrt{2 E_n x} \left(B + 2 \pi n\; {\cot} \frac{z}{2}\right).
\ee
Thus, all the eigenfunctions of $H$ exhibit a scaling behaviour,
i.e. $\psi_n \sim \sqrt{x}$, in the near-horizon region $\Delta$.
It should be stressed that this analysis is made possible
by utilizing the freedom in the choice of $z$.
The parameter $z$, which labels the self-adjoint extensions
of $H$, thus plays a crucial role in
establishing the self-consistency of this analysis.

We conclude this section with the following remarks:

\noindent
1. A particular choice of $z$ is
equivalent to a choice of domain for the differential operator $H$.
Physically,
the domain of an operator is specified by boundary conditions. A specific
value of $z$ is thus directly related to a specific choice of boundary
conditions for $H$. Thus,
we see that the system exhibits non-trivial scaling behaviour
only for a certain class of boundary
conditions.
These boundary conditions play a conceptually
similar role to the fall-off 
conditions as discussed in Ref. \cite{carlip,solo}.

\noindent
2. The analysis above provides a qualitative argument which suggests that
the scaling behaviour in the presence of a black hole
should be observed within a region
$\Delta$. Although $\Delta$ belongs to the near-horizon region
of the black hole, it
does not actually contain the event horizon. Our picture is thus similar in
spirit to the stretched horizon scenario of Ref. \cite{suss}.

\sxn{Density of States and Entropy}

In the analysis presented above, the information about the spectrum of the
Hamiltonian in the Schwarzschild 
background is coded in the parameter $z$. The wavefunctions and 
the energies of the bound states depends smoothly on $z$. Thus, within the
near-horizon region $\Delta $, we propose to identify
\be
{\tilde \rho} (z) \equiv \sum^{\infty}_{n=0} |\psi_n (z)|^2 
\ee
as the density of states for this system written in terms of the variable
$z$. ${\tilde \rho} (z) dz$ 
counts the number of states when the self-adjoint parameter lies between $z$
and $z + dz$. As mentioned before, within the region $\Delta $ 
$z$ is positive and
satisfies the consistency condition $z \sim 0$. From Eqns. (4.5)
and (5.4) we therefore see that the term
with $n=0$ provides the dominant contribution 
to the sum in Eqn. (6.1). The contribution of the terms with 
$n \neq 0$ to the
sum in Eqn. (6.1) is exponentially small for large ${\rm cot} \frac{z}{2}$.
Physically this
implies that the capture of the minimal probe excites only the lowest energy 
state in the near-horizon region of the massive black hole. 
The density of energy states of the black hole in the region $\Delta$ can
therefore be written as
\be
{\tilde \rho} (z)  \approx | \psi_0 |^2 = 
2 B^2 ~ e^{ {{ \pi} \over  4} {\rm cot} {z \over 2}}.
\ee
As mentioned before, within the region $\Delta$, ${\rm cot} {z \over 2}$
is a large and positive number. We thus find that 
the density of states of a massive black hole
is very large in the near-horizon region.

In order to proceed, we shall first provide a physical interpretation of the
self-adjoint parameter $z$ using the Bekenstein-Hawking entropy formula. 
To this end, recall that in our formalism, the capture of the scalar field
probe gives rise to the excitations of the black hole which 
subsequently decay by emitting Hawking radiation. 
A method of deriving density of states and entropy 
for a black hole in a similar physical setting using quantum mechanical
scattering theory
has been suggested by 't Hooft \cite{hooft}. This simple
and robust derivation uses the black hole
mass and the Hawking temperature as the only physical inputs and is
independent of the microscopic details of the system.
The interaction of infalling matter
with the black hole is assumed to be described by Schrodinger's
equation and the relevant emission and absorption cross sections 
are calculated using Fermi's Golden Rule. Finally, 
time reversal invariance (which is equivalent to CPT invariance in
this case) is used to relate the emission and absorption cross sections.
The density of states for a massive black hole of mass $M$ obtained
from this scattering calculation is given by 
\be
\rho (M) = e^{4 \pi M^2 + C^{\prime}} = e^S,
\ee
where $C^{\prime}$ is a constant and $S$ is the black hole entropy. 
It may be noted
that for the purpose of deriving Eqn. (6.3) the infalling matter was
described as particles. However, 
the above derivation of the density of states is independent of the
microscopic details and is valid for a general class of infalling matter. 

We are now ready to provide a physical interpretation of the parameter $z$. 
First note that the density of states calculated in Eqns. (6.2) and (6.3)
correspond to the same physical situation described in terms of different
variables. In our picture, the near horizon dynamics of the scalar field
probe contains information regarding the black hole
background through the self-adjoint parameter $z$. The same information
in the formalism of 't Hooft is contained in the black hole mass $M$. It is
thus meaningful to relate the density of states 
in our framework (cf. Eqn. 6.2) to that given by Eqn. (6.3). If these
expression describe the same physical situation, we are led to the
identification
\be
\frac{ \pi }{4} {\rm cot} {z \over 2} = 4 \pi  M^2.
\ee
Note that the analysis presented here is valid only for massive black holes. 
We have also seen that in the near-horizon region $z$ must be positive and 
obey a consistency condition such that
that ${\rm cot} {z \over 2}$ is large positive number. Thus the relation
between $z$ and $M$ given by Eqn. (6.4) is consistent with the constraints
of our formalism.
We therefore conclude that the self-adjoint parameter $z$ has a physical
interpretation in terms of the mass of the black hole.

As stated above, the density of states of the black hole can be expressed 
either in terms of the variable $z$ or in terms of $M$. In view of the
relation between $z$ and $M$, we can write 
\be
{\tilde \rho} (z) dz \sim |J| \rho (M) dM,
\ee
where $J = \frac{d z}{d M}$ is the Jacobian of the transformation from the
variable $z$ to $M$. When $z \sim 0$, using Eqns. (6.3) and (6.4) we get
\be
{\tilde \rho} (z) dz \sim e^{4 \pi M^2} \frac{1}{M^3} dM 
\sim e^{4 \pi M^2 - \frac{3}{2} {\rm ln} M^2} dM .
\ee
The presence of the logarithmic correction term in the above equation is
thus due to the effect of the Jacobian.

Finally, the entropy for the Schwarzschild black hole obtained from 
Eqn. (6.6) can be written as 
\be
S =  S(0) - \frac{3}{2} {\rm ln} S(0) + C ,
\ee
where $ S(0) = 4 \pi M^2$ is the Bekenstein-Hawking entropy and 
$C$ is a 
constant. Thus the leading correction to the Bekenstein-Hawking entropy is 
provided by the logarithmic term in Eqn. (6.7) with a coefficient of
$- \frac{3}{2} $. The subleading corrections to the entropy coming from the 
$n \neq 0 $ terms of Eqn. (6.1) are 
exponentially suppressed. 
As stated before, this is precisely the structure associated with the
expression of black hole entropy whenever the same is calculated within a
conformal field theoretic formalism \cite{danny}. 
We are thus led to conclude that the expression of the black hole entropy
obtained in our formalism provides a strong indication of
the underlying near horizon-conformal structure present in the system.

\sxn{Conclusion}

In this paper, we have analyzed the near-horizon properties of the
Schwarzschild black hole, using a  scalar field as a
simple probe of the system. We restricted attention 
to the time-independent modes of the scalar
field, and this allowed us
to obtain a number of interesting results
regarding the near-horizon properties of the black hole.
It is possible
that more sophisticated probes of general field
configurations may lead to additional information.

The factorization of $H$,
leading to the algebraic formulation of section 2,
is a process which appears to be essentially classical.
As discussed, the algebra appearing in Eqns. (2.11-2.15) does not at
first contain all the Virasoro generators.
The requirement of unitarity of the
representation leads to the inclusion of all the generators.
It is thus fair to say that the full Virasoro algebra appears in our
framework only at the
quantum level. The operator $H$ does not belong to ${\cal {A}}$ but
is contained in the enveloping algebra of the Virasoro generators. 
The enveloping algebra is the natural tool that 
is used to obtain representations of ${\cal {A}}$. Thus, even
though $H$ is not an element of ${\cal {A}}$, it nevertheless
has a well-defined
action in any representation of ${\cal {A}}$. It is this feature
that makes the algebraic description useful.

In section 3,
we summarized some results from the representation theory of
${\cal {A}}$. The operator $H$ is now treated at the quantum level,
and the corresponding eigenvalue problem is studied
using the representations
of ${\cal {A}}$. Unitarity again plays a role in restricting the space of
allowed representations. It is interesting to note that for all values of
the coupling $a \geq - \frac{1}{4}$, the value of the central charge in the
representation of ${\cal {A}}$ is equal to 1 \footnote{It may be noted that
the representation space for the $c=1$ conformal field theory
is spanned by tensor densities of weight
$\frac{1}{2}$, i.e. spin $\frac{1}{2}$ \cite{kac}.}.
Other black holes
which have $a$ in this range would exhibit a universality in this
regard. As mentioned in section 3, the relationship of $c$ calculated here
to that appearing in the entropy calculation of a particular black hole   
would depend
on other factors which are likely to break the universality.

If a Virasoro algebra is associated with the near-horizon dynamics,
then some reflection of it should appear in the spectrum of $H$. In
particular, we can expect that the wave functions of $H$ in the
near-horizon region should exhibit scaling behaviour.
Such a property was indeed
found in a band-like region near the horizon.
It is interesting to note that
this band excludes the actual horizon. This is similar in spirit to the
stretched horizon scenario of black hole dynamics.
The parameter $z$ describing the
self-adjoint extensions of $H$ is restricted to a set of values in this
process. This implies that the near-horizon wave functions
exhibit scaling behaviour only for a certain class of boundary
conditions.
It is important to note that boundary
conditions also played a crucial role in proving the existence of a Virasoro
algebra in Ref. \cite{carlip, solo}.
This feature provides a common thread in these
different approaches towards the problem.

In order to investigate this idea further, we first observe that the 
self-adjoint
parameter $z$ describes the domain of the Hamiltonian and directly
determines the spectrum within this formalism. On the other hand, the black
hole mass $M$ must also play a role in determining the spectrum. 
These two parameters thus play a conceptually similar role and it is 
expected that they will be related. 

The next step of our 
analysis was based on the identification of these bound states with the
excitations of the black hole resulting from the capture of the scalar field
probe. 
These excitation would eventually decay through the emission of
Hawking radiation. This process is described by 
quantum mechanical
scattering theory in terms of the density of states of the black hole
which is a function of the variable $M$ \cite{hooft}.
On the other hand, the
density of states following from our formalism is a smooth
function of $z$. Identifying 
these two expressions of the density of states 
leads to a quantitative relation between $z$ and $M$ which is
consistent with the constraints of the system. Such a relation also
provides a physical interpretation of the self-adjoint parameter in terms of
the mass of the black hole.

The relation between $z$ and $M$ naturally leads to the identification of
black hole entropy within this formalism. The entropy thus obtained 
contains the usual Bekenstein-Hawking term together with 
a leading logarithmic correction which has  
$- \frac{3}{2}$ as the coefficient. Moreover, 
the subleading non-constant corrections are shown to
be exponentially suppressed. It has been observed that 
the expression for the black hole entropy is expected to have precisely this
structure whenever it is calculated within the conformal field 
theoretic formalism \cite{danny} and possibly even for the non-unitary case
\cite{moore}.
Thus the expression that we obtain for the 
black hole entropy provides strong support to the hypothesis 
of an underlying conformal structure in the near-horizon region of the 
Schwarzschild black hole.

It is known that the near-horizon dynamics of various black holes
is described by an operator of the form $H$
\cite{trg,town1}, for different values of $a \geq -\frac{1}{4}$.
Any such operator can be factorised as
in Eqn. (2.2) and the above analysis will also apply
to these black holes.
It has been claimed  
in \cite{carlip, solo} that a Virasoro algebra is associated with a
large class of black holes in arbitrary dimensions.  
It seems plausible that the near-horizon dynamics    
of probes in the background of these black holes     
would be described by an operator of the form of $H$ and the method 
presented here
can be used to analyze the entropy for such black holes as well.

\noindent
{\bf Acknowledgements}

K.S.G. would like to thank Danny Birmingham and Siddhartha Sen for
collaborations \cite{ksg1,ksg2} on which this talk is based.  

\bibliographystyle{unsrt}

\begin{thebibliography}{99}
\bibitem{strom} A. Strominger, JHEP {\bf 9802}, 009 (1998).
\bibitem{carlip} S. Carlip, Phys. Rev. Lett. {\bf 82}, 2828 (1999); Class.
Quant. Grav. {\bf 16}, 3327 (1999); Nucl. Phys. Proc. Suppl. {\bf 18}, 10 
(2000); gr-qc/0203001 and references therein. 
\bibitem{solo} S. N. Solodukhin, Phys. Lettt. {\bf B 454}, 213 (1999).
\bibitem{ghosh} O. Dreyer, A. Ghosh and J. Wisniewski, Class. Quant. Grav.
{\bf 18}, 1929 (2001).
\bibitem{brown} J. D. Brown and M. Henneaux, Commun. Math. Phys. {\bf 104} 
207 (1986).
\bibitem{thorn} K. S. Thorne, R. H. Price and D. A. Macdonald (Ed.), {\it  
Black Holes: The Membrane Paradigm}, (Yale University Press, London, 1986).
\bibitem{hooft1} G. 't Hooft, ``Dimensional Reduction in Quantum Gravity",    
Essay dedicated to Abdus Salam, gr-qc/9310026.
\bibitem{suss} L. Susskind, L. Thorlacius  
and J. Uglum, Phys. Rev. {\bf D 48}, 3743 (1993); L. Susskind, Phys. Rev.  
{\bf D 49}, 6606 (1994); L. Susskind, J. Math. Phys. {\bf 36}, 6377 (1995).
\bibitem{hooft} G. 't Hooft, Int. Jour. Mod. Phys {\bf A 11}, 4623 
(1996); G. 't Hooft, ``The
Holographic Principle", Talk at the International School of Subnuclear
Physics, Erice, 1999, hep-th/0003004.
\bibitem{trg} T. R. Govindarajan, V. Suneeta and S. Vaidya,
Nucl. Phys. {\bf B583}, 291 (2000).
\bibitem{fubini} V. de Alfaro, S. Fubini and G. Furlan, Nuovo Cim.
{\bf 34A}, 569 (1976).
\bibitem{vasiliev} E. Bergshoeff and M. Vasiliev, Int. Jour. Mod. Phys.
{\bf 10}, 3477 (1995).
\bibitem{town1} P. Claus, M, Derix, R. Kallosh, J. Kumar, P. K. Townsend and
A. V. Proeyen, Phys. Rev. Lett. {\bf 81}, 4553 (1998).
\bibitem{town2} G. W. Gibbons and P. K. Townsend, Phys. Lett. {\bf B454},
187 (1999).
\bibitem{ksg1} Danny. Birmingham, Kumar S. Gupta and Siddhartha Sen,
Phys. Lett. {\bf B505}, 191 (2001).
\bibitem{narn} H. Narnhofer, Acta Physica Austriaca {\bf 40}, 306 (1974).
\bibitem{reed} M. Reed and B. Simon, {\it Methods of Modern Mathematical
Physics}, volume 1, Academic Press, New York, 1972; volume 2, Academic  
Press, New York, 1975. 
\bibitem{ksg2} Kumar S. Gupta and Siddhartha Sen, Phys. Lett. {\bf B526},
121 (2002).
\bibitem{partha} Romesh K. Kaul and Parthasarathi Majumdar, Phys. Rev. Lett.
{\bf 84}, 5255 (2000).
\bibitem{car1} S. Carlip, Class. Quant. Grav. {\bf 17}, 4175 (2000).
\bibitem{trg1} T. R. Govindarajan, Romesh K. Kaul and V. Suneeta,  
Class. Quant. Grav. {\bf 18}, 2877 (2001).
\bibitem{das1} Saurya Das, Romesh K. Kaul and
Parthasarathi Majumdar, Phys. Rev. {\bf D 63}, 044019 (2001).
\bibitem{danny} Danny Birmingham and Siddhartha Sen, Phys. Rev. {\bf D 63}, 
04750 (2001).
\bibitem{partha1} Saurya Das, Parthasarathi Majumdar and Rajat K. Bhaduri,
hep-th/0111001.
\bibitem{rade} H. Rademacher, {\it Topics in Algebraic Number Theory},
Springer-Verlag, Berlin, 1973.
\bibitem{ram} G. H. Hardy and S. Ramanujan, Proc. Lond. Math. Soc. {\bf 2}, 
75 (1918).
\bibitem{case} K. M. Case, Phys. Rev. {\bf 80}, 797 (1950).
\bibitem{ksg} K. S. Gupta and S. G. Rajeev, Phys. Rev {\bf D 48}, 5940
(1993); H. E. Camblong, L. N. Epele, H. Fanchiotti and C. A. G. Canal,
Phys. Rev. Lett {\bf 85}, 1593 (2000).
\bibitem{godd} D. Friedan, Z. Qiu and S. Shenker, Phys. Rev. Lett {\bf 52},
1575 (1984); P. Goddard and D. Olive, Int. Jour. Mod. Phys. {\bf A1},
303 (1986).
\bibitem{kac} V. G. Kac and A. K. Raina, {\it Bombay Lectures on Highest 
Weight Representations of Infinite Dimensional Lie Algebras}, World 
Scientific, Singapore, 1987.
\bibitem{abr} M. Abramowitz and I.A. Stegun, {\em Handbook of
Mathematical Functions}, Dover, New York, 1970.
\bibitem{moore} J. A. Harvey, D. Kutasov, E. J. Martinec and G. Moore,
hep-th/0111154.
\end{thebibliography}

\end{document}